\documentclass{article}
\usepackage{graphicx} 
\usepackage{nopageno}
\usepackage{xcolor}
\usepackage{tikz}
\usepackage{graphicx}

\usepackage{amsmath}               
\usepackage{amsthm}                
\usepackage{amssymb}               
\usepackage{geometry}              
\usepackage{fancyhdr}              
\usepackage[shortlabels]{enumitem} 
\usepackage{enumitem}
\usepackage{mathrsfs}
\usepackage{mathtools}
\usepackage[ruled, linesnumbered]{algorithm2e}
\usepackage{tikz}
\usetikzlibrary{angles,quotes,positioning}
\tikzset{main node/.style={circle,fill=none,draw,minimum size=1cm,inner sep=0pt},}
\PassOptionsToPackage{hyphens}{url}\usepackage{hyperref}
\usepackage[autostyle=true,german=quotes]{csquotes}

\usepackage{bm}
\usepackage{bbm}
 
\usepackage{graphicx}
\usepackage{color}

\usepackage{scrextend}
\usepackage{multirow} 
\usepackage{authblk}
\usepackage{soul}

\setlist[enumerate]{leftmargin=*}  
\setlist[itemize]{leftmargin=*}        
\newlist{todolist}{itemize}{2}
\setlist[todolist]{label=$\square$}
\usepackage{pifont}

\newtheorem*{theorem*}{Theorem}

\newtheorem*{lemma*}{Lemma}

\newcommand{\p}[1]{\left( #1 \right)}

\newcommand{\cd}[0]{\cdot}

\usepackage{framed, listings}
\usepackage{thmtools,thm-restate}

\title{An Abundance of Katherines\footnote{With apologies to John Green \cite{green2006abundance}.}: \\The Game Theory of Baby Naming}

\author[1]{Katy Blumer} 
\author[1]{\\Kate Donahue}
\author[1]{\\Katie Fritz}
\author[1]{\\Kate Ivanovich}
\author[1]{\\Katherine Lee}
\author[1]{\\Katie Luo}
\author[1]{\\Cathy Meng}
\author[1]{\\Katie Van Koevering}

\affil[1]{Cornell University\footnote{Correspondence to Katy Blumer, Kate Donahue, Katie Van Koevering: \{keb297, kpd46, kav64\}@cornell.edu}}

\date{}

\begin{document}

\maketitle

\begin{abstract}
In this paper, we study the highly competitive arena of baby naming. Through making several Extremely Reasonable Assumptions (namely, that parents are myopic, perfectly knowledgeable agents who pick a name based solely on its \enquote{uniqueness}), we create a model which is not only tractable and clean, but also perfectly captures the real world. We then extend our investigation with numerical experiments, as well as analysis of large language model tools. We conclude by discussing avenues for future research. 
\end{abstract}

\section{Introduction}
The most important decision in any child's life happens shortly after they are born and is made entirely without their input or approval - their naming. The name given to a child is traditionally kept throughout their life time and has significant impact on their future life path. This momentous decision is made by parents with little education in the game theory inherently present in the highly competitive field of naming. 

We attempt to assist these parents with a simple primer into the game theory underpinning the decision of naming. We will introduce the basic set-up of the naming game and formalize the parameters and incentives, wherein parents have some desired properties of the name. We then describe the pitfalls of the most simple interpretations of these models, particularly the dangers of myopic action. Finally, we present experimental results demonstrating the shift in name distributions under this model. These experiments underline the inherent risks in naming a child and highlight how altering various parameters can change the outcomes of naming strategies.

\section{Related works}

Surprisingly, no one has ever done any research on naming strategies (so long as you conveniently ignore \cite{acerbi2014biases, bentley2012accelerated,bush2018network,cila2019s,coulmont2016diffusion,glynn2002institutionalizing,golman2022hipsters,hahn2003drift,hanunderstanding,kessler2012you,lee2015theory,leonardelli2010optimal,li2012analyses,muller2020analyzing,newberry2022measuring,stoyneva2022demystifying,twenge2016still,varnum2011s} and likely other work).

\section{Model}
Naming a child is akin to choosing an outfit for the Oscars. It must be unique enough to stand out - no one wants to show up to the Oscars in the same dress - but it must also be similar enough to be recognizable as a name. Lady Gaga's meat dress is fine for an afternoon, but a child named \enquote{Meat Dress} would soon become discontented, if not the plaintiff of a lawsuit.  Thus, we model name selection based on the desired \enquote{uniqueness} of the name. 

\subsection{Name frequency and choice model}
First, we present our formal model. Assume there is some set of names $\mathcal{A}$. At a time point $i$, we assume there exists a discrete distribution of popularity over names $$f_i(a) = \nu$$ such that $f(a) \in [0, 1], \sum_{a \in \mathcal{A}}f_i(a) = 1$. For simplicity, we assume that every name has unique frequency: that is, $f_i(a_j) \ne f_i(a_k)$ for $j \ne k$. 

Next, we model parental preferences. It is well-known that parents are always in complete agreement over the name they would prefer to pick for their newborn child. Therefore, we will treat the parents of each child as a unit, and assume that each set of parents $j \in \mathcal{P}$ has some preference over the proportion of the population that would share the same name as their child (we also assume each parental unit has exactly 1 child). For example, $\mu_j = 0.01$ means that parental unit $j$ wants their child to have a name that is shared by 1\% of the population. We will use $$g(\mu) = p$$ to mean that $p \in [0, 1]$ proportion of parents want a name with popularity $\mu$. For example, if $\mu= 0.1$ and $p = 0.2$, then this means that 20\% of parents want a name with popularity 10\%. This set-up gives us convenient parameters for the model and just enough Greek letters to sound smart enough for publication.

In general, we will assume that parents are \emph{myopic}, with new parents having no concept of time but perfect access to baby name data. We find this a realistic assumption. Mathematically, parents at time step $i$ will pick the name $a$ that currently is closest to their desired frequency $\mu_j$.  
Given this assumption, then at time step $i+1$, the proportion of babies who have name $a$ is given by the total fraction of parents for whom name $a$ is closest to their desired frequency. 

\subsection{Satisfiability}
If parents are unable to infer the consequences of their actions and act myopically, then it can immediately be seen that some parents will be deeply unhappy with said consequences: for example, if $g(0.1) = 0.2$ (as in the example above), then the 20\% of parents who wished that their name has popularity 10\%, will end up with a name that is more popular than they anticipated (when $g(\mu)> \mu$). For instance, a parent might anticipate the name \enquote{Kate} would be a pleasantly traditional yet unique name with only moderate popularity. They would be wrong \cite{kat}. 

Conversely, if we had $g(0.1) = 0.05$ (or $g(\mu) < \mu$), then parents would end up with a name that is less popular than they anticipated. If $g(\mu) = \mu$, then we say the name with proportion $\mu$ is \emph{satisfied}. Ideally, we would like every name to be satisfied, or $g(\mu) = \mu$ for all $\mu \in [0, 1]$. However, that would give us a distribution with total probability $>1$, which is the sort of thing that makes statisticians sad. Instead, we see that the entire naming distribution $g(\cd)$ is satisfied if every name with \emph{nonzero popularity} $\mu>0$ has exactly $\mu$ fraction of the population that desires this name: 
$$\begin{cases}
    g(\mu) = \mu &  \mu > 0\\
    g(\mu) = 0  & \text{otherwise}
\end{cases}$$

\subsection{Stability}
Next, we consider an alternative property that we may wish to have: that the distribution of names be \emph{stable}. If an arrangement is stable, this means that given an existing distribution $f_i(a)$ and a parental preference distribution $g(\mu)$, every name's frequency will be exactly the same at the next time step $i+1$: 

$$f_{i+1}(a) =  f_i(a) \quad \forall a \in \mathcal{A}$$

The simplest way to achieve stability is if every parent would prefer to name their child after themselves. That is, if every parent wishes their child to have the same \enquote{uniqueness} of naming they do, a sort of inheritability of uniqueness - we name this the Dweezil Principle.

\subsection{Extremely Reasonable Assumptions}
The above model contains several Extremely Reasonable Assumptions (ERAs). The first ERA is the very conservative assumption that there is only one gender, with all children and all names adhering to the same gender. Thus any child may be given any name, so long as it exists in the names list\footnote{If a fixed names list is good enough for the Scandinavians, it's good enough for us \cite{preapproved_names}}. Another ERA is the Mayfly Parenthood Assumption, in which all parents perish immediately upon naming their child, which makes the math substantially easier.

\section{Illustrative example: power law distribution} \label{sec:power-law}
In this section, we consider the case where both $f(\cd)$ and $g(\cd)$ are given by \emph{power law distributions}.

\begin{figure}[h]
    \centering
    \includegraphics[width=1\linewidth]{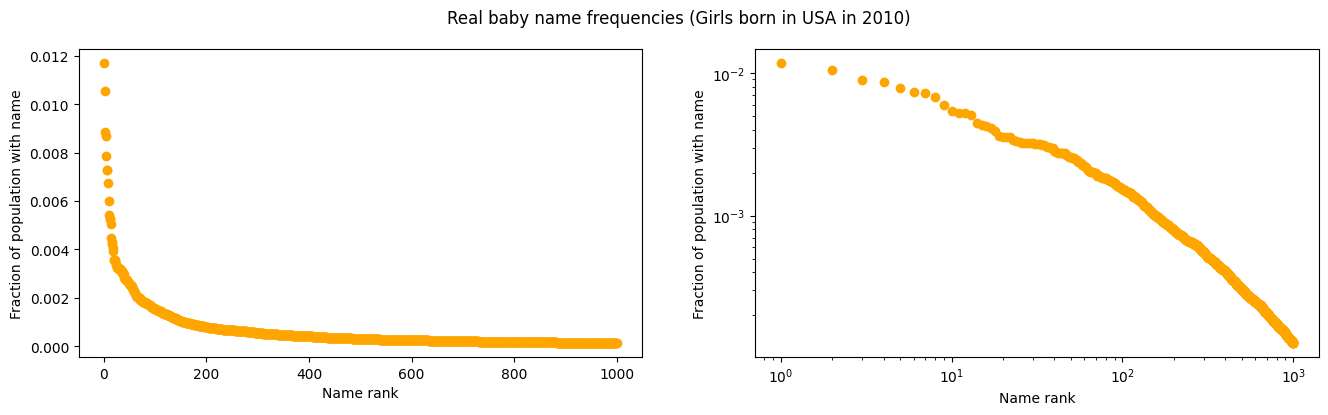}
    \caption{Name frequencies from the Social Security Administration, for girls born in 2010\cite{ssa}. Note the rough power-law shape.}
    \label{fig:real-data}
\end{figure}

\subsection{Modeling $f(a), g(\mu)$}
We begin by defining our variable for name popularity. The popularity of names has been shown to follow a power law distribution \cite{li2012analyses}  (see Figure \ref{fig:real-data}) . We can model this as: 
$$f(a) = K\cdot a^{-t}$$
where $a$ denotes the rAnk of the name within $a \in [1, N]$, $t$ is another constanT, and $K$ is a normalization konstant.

Next, we consider how parents pick the uniqueness of names for their children: the function $g(\cd)$. Because we are in the Power Law subsection, we will also assume that this distribution of parent preferences is a power law. 
To minimize our use of variables, we will model this as:
$$g(a) = K'\cd \p{a'}^{-t'} $$
where $a'$ is the desired frequency of the name (within the range $[\epsilon, 1]$, for $\epsilon>0$), $t'$ is another constant, and $K'$ is a normalization constant.

Note that for $t'>0$, we have that $g(\mu)$ is decreasing in $\mu$: that is, more parents prefer names that are \emph{uncommon}. Conversely, for $t'<0$, we have that $g(\mu)$ is increasing in $\mu$: parents prefer names that are \emph{common}. 

\begin{figure}
\centering 
    \includegraphics[width=4in]{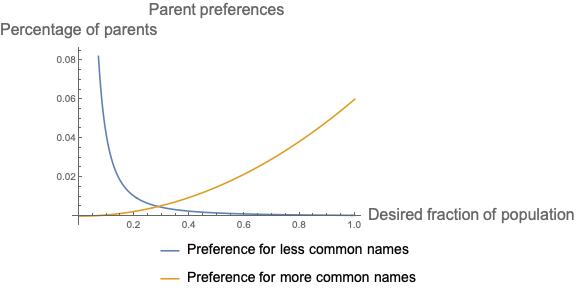}
    \caption{Examples of different parental preferences: a preference for less common names (blue) and  preference for more common names (orange).}
    \label{fig:gpower}
\end{figure}

\subsection{Picking names: stability}

Given a $f_i(a)$ and $g(\mu)$, the distribution $f_{i+1}(a)$ at the next time step is given by: 

$$f_{i+1}(a) = g(f_i(a)) = K' \cd \p{K \cd a^{-t}}^{-t'} = K' \cd K^{-t'} \cd a^{t \cd t'}$$
Note that this is again a power law distribution\footnote{This is why we like power laws.}, with parameter $t \cd t'$. Careful selection of the parameter $\epsilon$ and analysis of the constants $K, K'$ can confirm that this is a valid probability distribution, a task which we \st{couldn't be bothered to do} leave for the interested reader. Next, we can analyze the properties of the resulting distribution.

\subsubsection{Picking an uncommon name: a futile quest}

\begin{figure}
\centering 
    \includegraphics[width=3in]{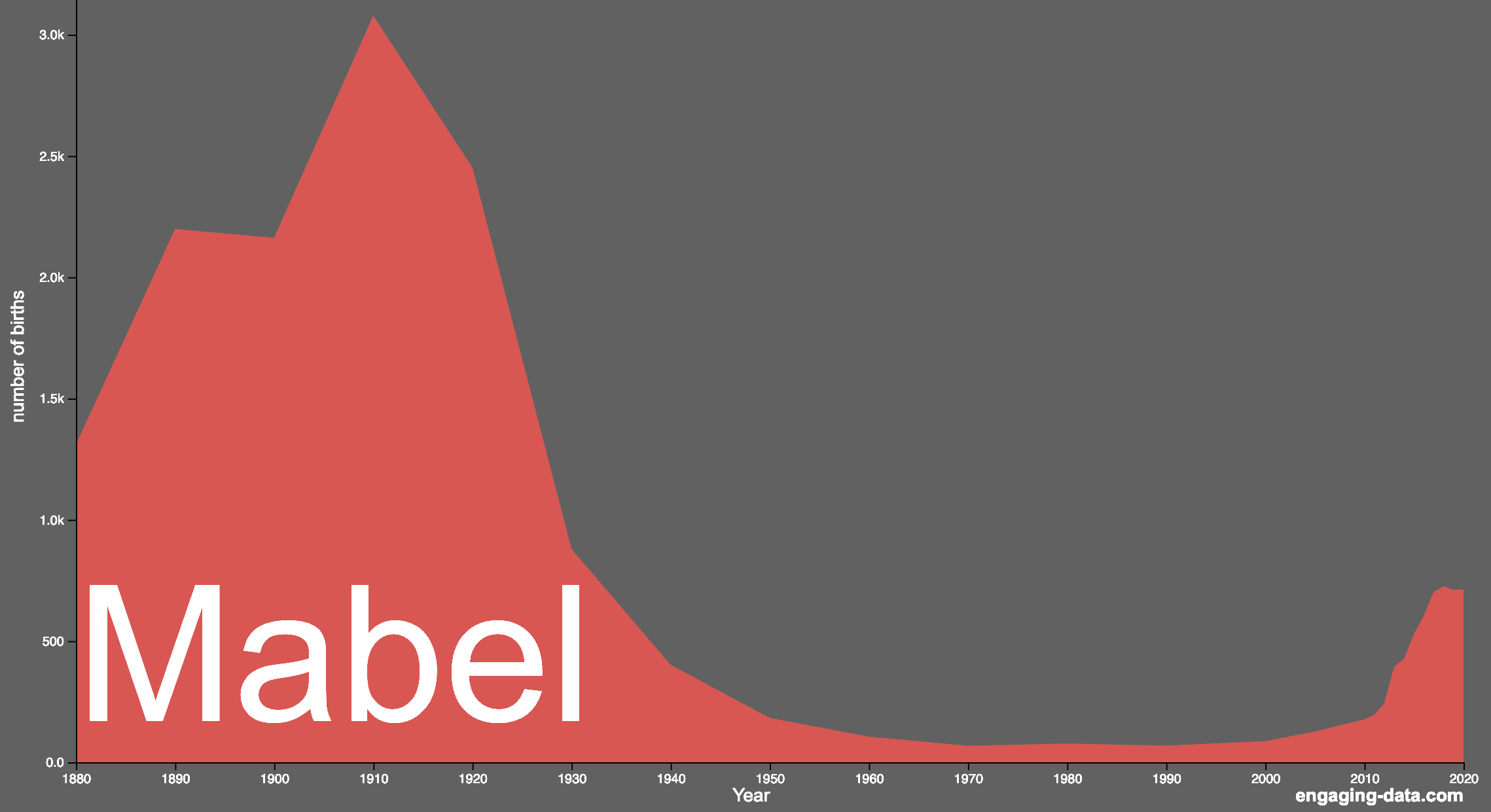}
    \caption{Frequency of name Mabel (image from \cite{mabel_image_source}).}
    \label{fig:mabel}
\end{figure}

First, we consider the case where $t'$ is positive, which corresponds to the case where most parents prefer uncommon names. In this setting, we know that $t\cd t'$ is also positive, which means that $f_{i+1}(a)$ is \emph{increasing} in $a$. This tells us that a name with high popularity at time $i$ will have \emph{low} popularity at time $i+1$. 
Given an original distribution of frequency over names shown in blue in Figure \ref{fig:shift}, if parents have a preference for less common names, the resulting distribution will look like the orange curve: the least popular name has suddenly become extremely popular, and what was popular at time step $i$ has become horribly pass\'e by time step $i+1$. This means that names will see-saw in popularity from one time step to another, which explains why both your grandmother and niece are named Mabel (Figure \ref{fig:mabel}). 

\subsubsection{Picking a common name: naming event horizon}

\begin{figure}
\centering 
    \includegraphics[width=4in]{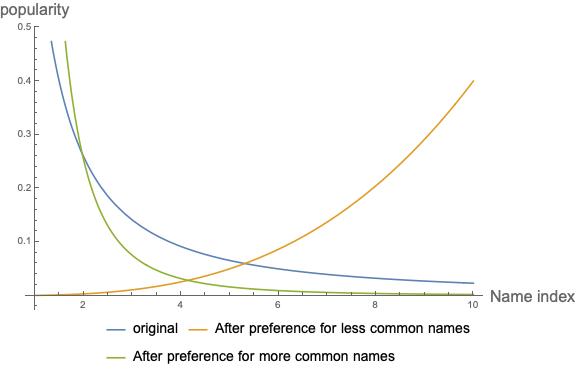}
    \caption{Examples of naming frequency at time step $i+1$: the original distribution (blue), the distribution after a preference for less common names (orange), and a preference for more common names (green).}
    \label{fig:shift}
\end{figure}

However, parents might believe there are benefits to their child sharing a name with others (e.g., see \cite{kat}). We can model this with the case where $t'$ is negative, which means that most parents would prefer names that are relatively popular, with only a few parents preferring names that are less common. Mathematically, this means that $t \cd t'$ is also negative, which means that $f_{i+1}(a)$ is \emph{decreasing} in $a$. This means that popular names at time step $i$ are also popular at time step $i+1$: the relative order of name popularity stays the same (as shown in the green curve in Figure \ref{fig:shift}). 

If this process is repeated $n$ times, with the same parental preferences, then the resulting power law distribution would have exponent $t \cd t'^{n}$. For $t'>1$ (the \enquote{event horizon}), the means that the most popular names gobble up almost all of the population, resulting in a black hole of names wherein infinite density will eventually fall upon just one name.

\section{Simulations}
\begin{figure}[h]
    \centering
    \includegraphics[width=1\linewidth]{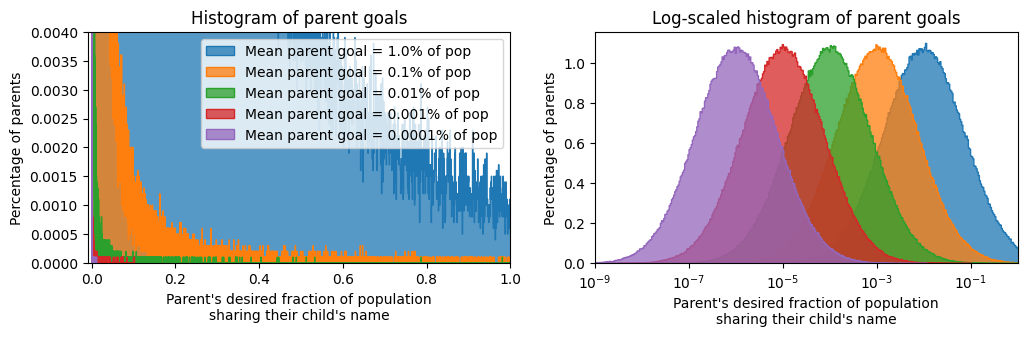}
    \caption{Parent preference distributions used in experiments. Logarithmic x-axis on the right, showing the log-normal distribution shape. Each histogram represents a sample of 1 million parents from the given distribution.}
    \label{fig:exp-parent-hist}
\end{figure}

\begin{figure}[h]
    \centering
    \includegraphics[width=0.5\linewidth]{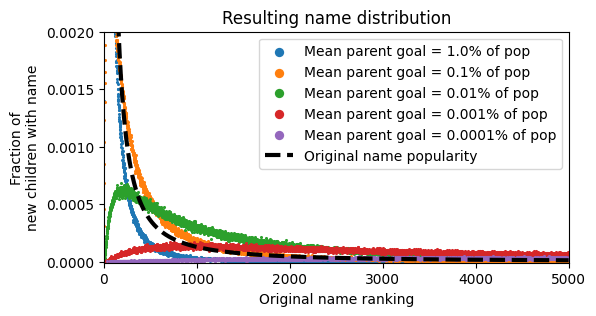}
    \caption{Resulting name distributions for children of parents with given preference distribution.}
    \label{fig:exp-name-dist}
\end{figure}

\begin{figure}[h!]
    \centering
    \includegraphics[width=1\linewidth]{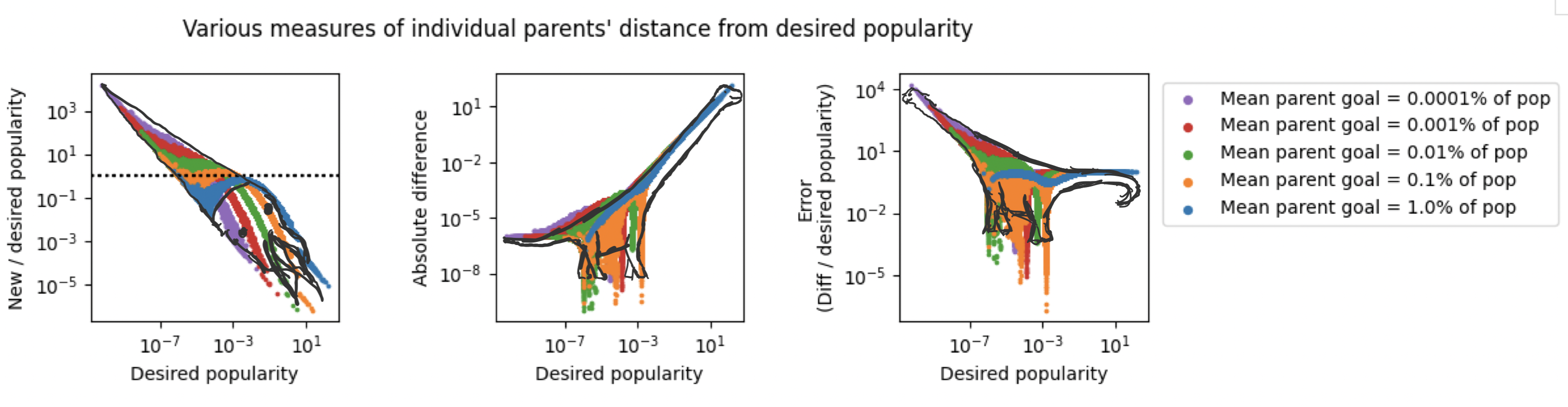}
    \caption{Various measures of parent error, meaning the difference between their chosen name's popularity and their desired name popularity. From left to right: ratio (true popularity divided by desired popularity), absolute difference (absolute value of true$-$desired popularity), and error (absolute difference divided by desired popularity). Note the adorable dinosaurs.}
    \label{fig:exp-dinosaurs}
\end{figure}

For simulations we use a log-normal distribution of parent preferences, rather than a power law as in Section \ref{sec:power-law}, because a certain author was having issues with SciPy. We baselessly claim a log-normal makes sense because name \enquote{uniqueness} is logarithmic; that is, a name belonging to 0.01\% of the population is roughly twice as unique as a name belonging to 0.1\% of the population (when comparing to a baseline name with 1\% popularity). 

We work with five different parent preference distributions, with popularity modes of 1\% to 0.0001\% (Figure \ref{fig:exp-parent-hist}). From each preference distribution, we generate a sample of 1 million parents, who each choose the name which is currently closest to their desired popularity. The resulting distribution of names is shown in Figure \ref{fig:exp-name-dist}.  If parents prefer more unique names, the name distribution flattens, since there are many more names at the low-popularity end of the scale for parents to divide between. If parents prefer very popular names, the distribution is even more heavily weighted at the popular end than the original power law distribution. The preference distribution centered at 0.1\% in fact closely resembles the original power law distribution.

Measures of parent error (distance from goal popularity) do not show particularly interesting patterns, but are included because they look like dinosaurs (Fig. \ref{fig:exp-dinosaurs}).

\section{Obligatory Kat-GPT experiment}

\begin{figure}[h]
\centering 
    \includegraphics[width=3in]{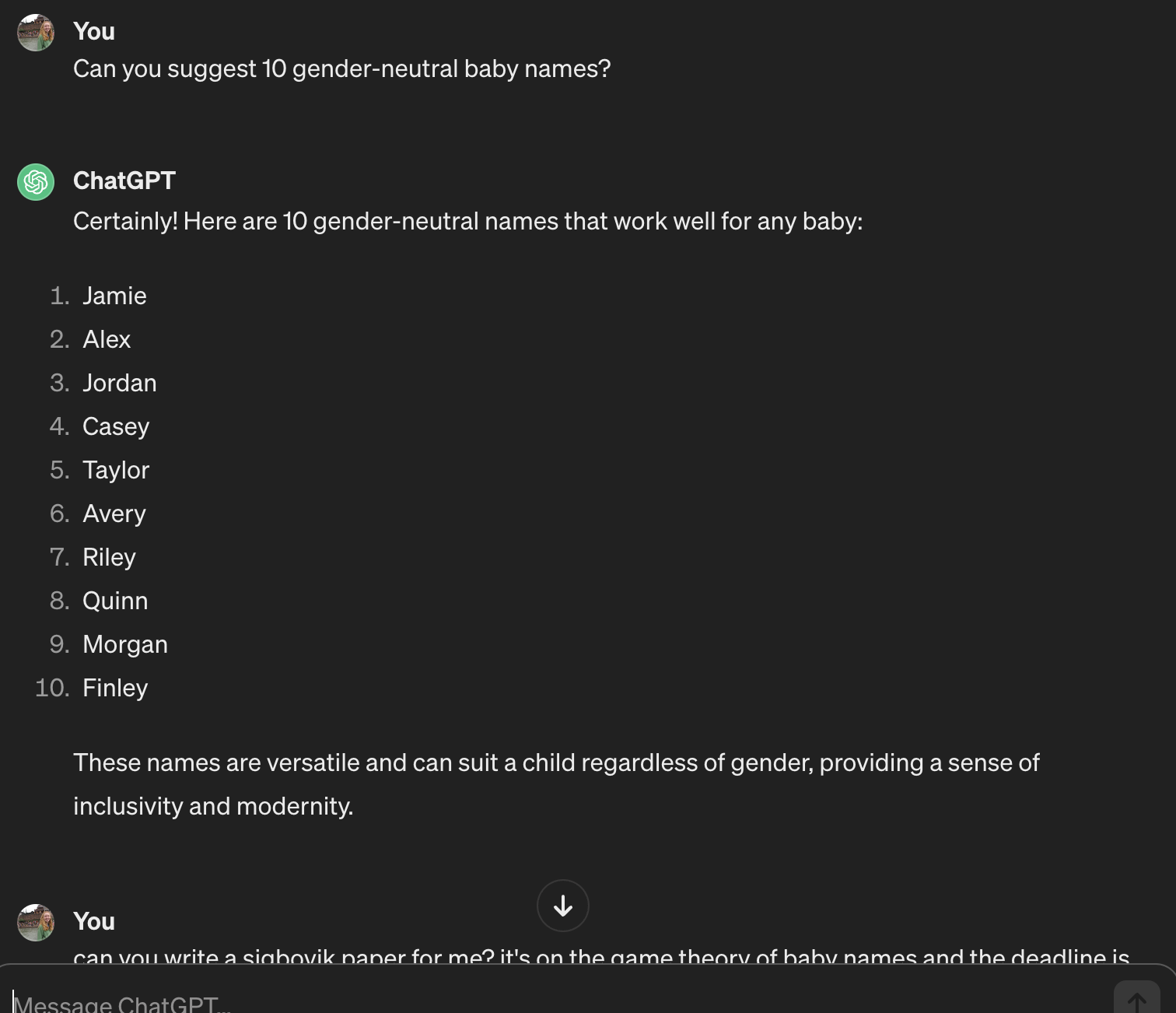}
    \caption{Experiments with Kat-GPT}
    \label{fig:katgpt}
\end{figure}

Because this paper was written in 2024, we include an obligatory section involving generative AI and LLMs. It is fortunate that the most popular LLM of the year appears to be custom-made for this experiment: Kat-GPT. Specifically, we asked Kat-GPT to give us its top ten names for a) a girl, b) a boy, and c) a gender-neutral name (see Figure \ref{fig:katgpt}). Then, we calculated the frequency of those names within their gender category (or within the total population, for gender-neutral). The results are given within Table \ref{tab:namepop}. 

The difference in mean popularity between \enquote{girl} and \enquote{gender-neutral} names and between \enquote{boy} and \enquote{gender-neutral} names are both statistically significant at the $p\leq 0.05$ level, while the difference in mean popularity between \enquote{girl} and \enquote{boy} names is statistically significant at the $p\geq 0.05$ level. 

\begin{table}[h!]
\centering 
\begin{tabular}{|c|c|c|c|}
\hline
                  & \enquote{girl} names & \enquote{boy} names & \enquote{gender-neutral} names \\ \hline
Mean popularity   & 0.6728               & 0.666               & 0.126                          \\ \hline
Std of popularity & 0.134                & 0.212               & 0.0619                         \\ \hline
\end{tabular}
\caption{Popularity of names given by ChatGPT ($N=10^{-e^{\pi\cd i}}$ queries for each category)}
\label{tab:namepop}
\end{table}

\section{Extensions \& Future Work}
In this section we include multiple extensions that we considered but ultimately were too lazy to actually finish. 
\subsection{Creation of new names}
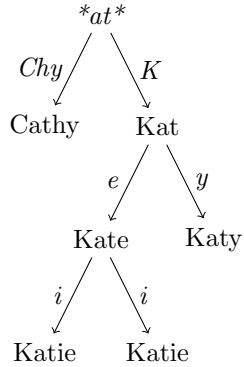
\begin{figure}[h]
    \centering
\begin{tikzpicture}
    \node {\emph{*at*}}
    child { node {Cathy} edge from parent [->] node [left] {\emph{Chy}} }
    child { node {Kat} 
        child { node {Kate}
        child {node {Katie}
        edge from parent [->] node [left] {\emph{i}}
        }
        child {node {Katie}
        edge from parent [->] node [right] {\emph{i}}
        }
        edge from parent [->] node [left] {\emph{e}}
        }
        child { node {Katy}  
        edge from parent [->] node [right] {\emph{y}}
        }
        edge from parent [->] node [right] {\emph{K}}
    }
    ;
\end{tikzpicture}
    \caption{An example of name mutation, where multiple novel (but related) names are displayed with their edit distance from each other.}
    \label{fig:Kattree}
\end{figure}
One of our Extremely Reasonable Assumptions is that there was a fixed list of names. However, new names have occasionally been documented in the wild \cite{okasha2016women}. One extension could consider a strategy whereby parents could pick a name $a'$ that has some distance $d(a, a')$ from an established name, and derive cost relating both to the popularity of the \enquote{base name} $a$ and the distance $d(a, a')$. 

$$\min_{a, a' \in \mathcal{A}}\left|K \cd {a'}^{-t} - \mu_i\right| + \lambda \cd d(a, a')$$
where $d(a, a')$ is a distance metric between names (e.g., see Figure \ref{fig:Kattree}).


\subsection{Non-myopic parents}
Another one of our Extremely Reasonable Assumptions was that parents are myopic: that is, if they have a desired name frequency $a'$, they simply blindly pick the name that currently has frequency $a'$, which can lead to over- or under-shooting their desired frequency. However, emerging research \cite{lasker1932lasker} suggests that people in fact reason strategically about their actions; if this is borne out, considering non-myopic parents may be an interesting avenue of future research (i.e. not us). 

\section{Conclusions and implications}
The science of naming has a long and illustrious history that we didn't bother to look at. Instead, we arbitrarily assigned a new(?) model to describe how parents ought to name their children - namely probabilistically.  This model has interesting implications, most interestingly that all naming strategies are futile. 
We also print some plots, for both educational and entertainment purposes, which further emphasize these points and have some nice dinosaurs. But overall, we find only one rule really matters when naming a child: when in doubt, name it Kate.

\section{Acknowledgements}
We thank F.D.C. Willard for helpful discussions. Simon Shindler contributed significantly to the aesthetic of Figure \ref{fig:exp-dinosaurs}, but could not be named an author for obvious reasons.

We also want to thank the many dozens of people who have confused us for one another at conferences. We enjoyed meeting you.

Finally, we thank Katie's mom for editing this paper and the rest of the parents for making the provably optimal (see above) name choice.


\bibliographystyle{acm}

\bibliography{works_cited}

\begin{thebibliography}{10}

\bibitem{preapproved_names}
Some countries have a list of preapproved baby names.
\newblock {\em Interesting Facts\/}.

\bibitem{ssa}
Popular baby names, 2024.
\newblock {https://www.ssa.gov/OACT/babynames/index.html}.

\bibitem{mabel_image_source}
{US} baby name popularity visualizer, 2024.
\newblock {https://engaging-data.com/baby-name-visualizer/}.

\bibitem{acerbi2014biases}
{\sc Acerbi, A., and Bentley, R.~A.}
\newblock Biases in cultural transmission shape the turnover of popular traits.
\newblock {\em Evolution and Human Behavior 35}, 3 (2014), 228--236.

\bibitem{bentley2012accelerated}
{\sc Bentley, R.~A., and Ormerod, P.}
\newblock Accelerated innovation and increased spatial diversity of us popular culture.
\newblock {\em Advances in Complex Systems 15}, 01n02 (2012), 1150011.

\bibitem{kat}
{\sc Blumer, K., Donahue, K., Fritz, K., Ivanovich, K., Lee, K., Luo, K., Meng, C., and Van~Koevering, K.}
\newblock An abundance of {Katherines}: The game theory of baby naming.
\newblock {\em SIGBOVIK\/} (2024).

\bibitem{bush2018network}
{\sc Bush, S.~J., Powell-Smith, A., and Freeman, T.~C.}
\newblock Network analysis of the social and demographic influences on name choice within the uk (1838-2016).
\newblock {\em PLoS One 13}, 10 (2018), e0205759.

\bibitem{cila2019s}
{\sc Cila, J., and Lalonde, R.~N.}
\newblock What’s in a name? motivations for baby-naming in multicultural contexts.
\newblock {\em Contemporary Language Motivation Theory: 60 Years Since Gardner and Lambert (1959) 3\/} (2019), 130.

\bibitem{coulmont2016diffusion}
{\sc Coulmont, B., Supervie, V., and Breban, R.}
\newblock The diffusion dynamics of choice: From durable goods markets to fashion first names.
\newblock {\em Complexity 21}, S1 (2016), 362--369.

\bibitem{glynn2002institutionalizing}
{\sc Glynn, M.~A., and Abzug, R.}
\newblock Institutionalizing identity: Symbolic isomorphism and organizational names.
\newblock {\em Academy of Management journal 45}, 1 (2002), 267--280.

\bibitem{golman2022hipsters}
{\sc Golman, R., Bugbee, E.~H., Jain, A., and Saraf, S.}
\newblock Hipsters and the cool: A game theoretic analysis of identity expression, trends, and fads.
\newblock {\em Psychological review 129}, 1 (2022), 4.

\bibitem{green2006abundance}
{\sc Green, J.}
\newblock {\em An Abundance of Katherines}.
\newblock Penguin Young Readers Group, 2006.

\bibitem{hahn2003drift}
{\sc Hahn, M.~W., and Bentley, R.~A.}
\newblock Drift as a mechanism for cultural change: an example from baby names.
\newblock {\em Proceedings of the Royal Society of London. Series B: Biological Sciences 270}, suppl\_1 (2003), S120--S123.

\bibitem{hanunderstanding}
{\sc Han-Wu-Shuang, B., Hua-Jian, C., and Yi-Ming, J.}
\newblock Understanding the rise of unique names: The emphasis on uniqueness matters.

\bibitem{kessler2012you}
{\sc Kessler, D.~A., Maruvka, Y.~E., Ouren, J., and Shnerb, N.~M.}
\newblock You name it--how memory and delay govern first name dynamics.
\newblock {\em PloS one 7}, 6 (2012), e38790.

\bibitem{lasker1932lasker}
{\sc Lasker, E.}
\newblock {\em Lasker's Manual of Chess: With 308 Diagrams}.
\newblock Printing-Craft, 1932.

\bibitem{lee2015theory}
{\sc Lee, M.~J., Do~Yi, S., Kim, B.~J., and Baek, S.~K.}
\newblock Theory of fads: Traveling-wave solution of evolutionary dynamics in a one-dimensional trait space.
\newblock {\em Physical Review E 91}, 1 (2015), 012815.

\bibitem{leonardelli2010optimal}
{\sc Leonardelli, G.~J., Pickett, C.~L., and Brewer, M.~B.}
\newblock Optimal distinctiveness theory: A framework for social identity, social cognition, and intergroup relations.
\newblock In {\em Advances in experimental social psychology}, vol.~43. Elsevier, 2010, pp.~63--113.

\bibitem{li2012analyses}
{\sc Li, W.}
\newblock Analyses of baby name popularity distribution in us for the last 131 years.
\newblock {\em Complexity 18}, 1 (2012), 44--50.

\bibitem{muller2020analyzing}
{\sc M{\"u}ller, J.}
\newblock {\em Analyzing given names: exploring potentials for a personalized name discovery on Nameling. net.}
\newblock PhD thesis, University of Kassel, Germany, 2020.

\bibitem{newberry2022measuring}
{\sc Newberry, M.~G., and Plotkin, J.~B.}
\newblock Measuring frequency-dependent selection in culture.
\newblock {\em Nature Human Behaviour 6}, 8 (2022), 1048--1055.

\bibitem{okasha2016women}
{\sc Okasha, E.}
\newblock {\em Women's names in Old English}.
\newblock Routledge, 2016.

\bibitem{stoyneva2022demystifying}
{\sc Stoyneva, I., and Vracheva, V.}
\newblock Demystifying entrepreneurial name choice: insights from the us biotech industry.
\newblock {\em New England Journal of Entrepreneurship 25}, 2 (2022), 121--143.

\bibitem{twenge2016still}
{\sc Twenge, J.~M., Dawson, L., and Campbell, W.~K.}
\newblock Still standing out: Children's names in the united states during the great recession and correlations with economic indicators.
\newblock {\em Journal of Applied Social Psychology 46}, 11 (2016), 663--670.

\bibitem{varnum2011s}
{\sc Varnum, M.~E., and Kitayama, S.}
\newblock What’s in a name? popular names are less common on frontiers.
\newblock {\em Psychological science 22}, 2 (2011), 176--183.

\end{thebibliography}
\end{document}